\documentstyle[preprint,revtex]{aps}

\newcommand\bd{\begin{displaymath}}
\newcommand\ed{\end{displaymath}}
\newcommand\be{\begin{equation}}
\newcommand\ee{\end{equation}}

\begin{document}
\begin{title}
{\bf THE EMERGENCE OF RADIATION \\
FROM GRAVITATIONAL POTENTIAL WELLS:\\
THE ABSENCE OF $\omega M$ EFFECTS}
\end{title}
\author{Richard H. Price and Jorge Pullin}
\begin{instit}
Department of Physics, University of Utah, Salt Lake City UT 84112
\end{instit}

\begin{abstract}

We consider a source of gravitational waves of frequency $\omega$,
located near the center of a massive galaxy of mass $M$ and radius
$R$, with $\omega\gg R^{-1}$. In the case of a perfect fluid galaxy,
and of odd-parity waves, there is no direct matter-wave interaction and
the propagation of the waves is affected by the galaxy only through
the curvature of the spacetime background through which the waves
propagate. We find that, in addition to the expected redshift of the
radiation emerging from the galaxy, there is a small amount of
backscatter, of order $M/\omega^2R^3$.  We show that there is no
suppression of radiative power by the factor $1+\omega^2M^2/4$ as has
been recently predicted by Kundu. The origin of Kundu's suppression
lies in the interpretation of a term in the expansion of the exterior
field of the galaxy in inverse powers of radius. It is shown why that
term is not related to the source strength or to the strength of the
emerging radiation.

\end{abstract}

\pacs{4.30.+x}

\section{Introduction}

If a source of gravitational radiation is located in or near a massive
body the curvature of spacetime caused by that body may influence the
generation and propagation of the radiation produced by the source.
If, for example, a quadrupole oscillator is located at the center of a
galaxy of mass $M$ and radius $R$, we might guess that the effect of
the surrounding galaxy on the radiation produced inside it is of the
same order, $M/R$, as the characteristic Newtonian potential, at least
in the case, $M/R\ll1$, that the galaxy is nearly Newtonian. (Here and
throughout we use units in which $G$ and $c$ are unity.) In this paper
we analyze just what these effects really are for high frequency waves.

To do this we consider a fairly specific astrophysical configuration.
We suppose that there is a source of gravitational radiation emitting
waves at frequency $\omega$, and confined to a central source region
$r<r_S$ of the galaxy. We require that the source be small compared to
the radius of the galaxy ($r_S\ll R$), We also require the source
frequency to be high enough so that there is a region, outside the
source ($r>r_S$) which is deep inside the galaxy ($r\ll R$) and in the
wave zone of the source ($r\gg\omega^{-1}$). In this region it is
meaningful to talk about the gravitational wave flux well inside the
galaxy.  For a typical galaxy $R\approx10^{23}$cm and $M\approx
10^{16}$cm, and for a kilohertz gravitational wave
$\omega\approx10^{-7}{\rm cm}^{-1}$.  For these values the region we
require is the range of radii satisfying ${\rm max}(r_S,10^{7}{\rm
cm})\ll r\ll10^{20}{\rm cm}$.

For such a configuration the standard analysis tells us that as the
high frequency waves propagate outward there are two effects of
spacetime curvature that affect their passage. First the frequency of
the waves is redshifted so that the frequency observed far outside the
galaxy is reduced from that at the source roughly by the factor
$(1+\Phi_0)$, where $\Phi_0$, the central potential, is of order
$M/R$.  The second effect is associated with the meaning of the radial
coordinate. If ``$r$'' is the usual (i.e., Schwarzschild) radial
coordinate, then for waves radially propagating outward, the rate
$dr/dt$ is slightly less than unity, and there is an attendant gradual
phase shift of the waves, of order $\omega M$, as viewed in the $r$
coordinate.  (For a discussion of propagation of gravitational waves,
and the distinction between generation and propagation for an
``isolated'' source, see Thorne\cite{KipRev}).

We consider what other effects influence the propagation of
gravitational waves, and find that there are interactions between the
spacetime curvature and the waves which are interesting as points of
principle, if not of astrophysical importance. There is, however, a
significant additional motivation for such a calculation, and a major
motivation for this paper.  Kundu\cite{Kundu,K2} has recently argued
that gravitational wave energy propagating out of a gravitational
potential well will be reduced in intensity by the factor
$(1+M^2\omega^2/4)^{-1}$.  Because $\omega M$ can be large (of order
$10^9$ for the typical numbers given above), such a reduction of
kilohertz gravitational wave signals originating in other galaxies
would make detection of signals impossible and would be of crucial
importance in connection with the detection of gravitational waves by
instruments now being developed.

The remainder of the paper is organized as follows: We start in
Sec.~II by outlining the mathematical origin of Kundu's argument that
gravitational radiation is suppressed. We then describe the argument
against suppression given by Kozameh, Newman, and Rovelli\cite{KNR},
and its relationship to the present work.  In Sec.~III we derive the
necessary connections between the NP formalism, used by Kundu, and the
formalism of metric perturbations. We show, in the Schwarzschild
exterior, how the NP projection $\Psi_0$ is related to the the Zerilli
function\cite{Zerilli} in the case of even-parity perturbations, and
in the odd-parity case to the function solving the Regge-Wheeler
equation\cite{RW}. For outgoing solutions of both parities,
suppression factors arise in the relationship between the terms
describing the strength of radiation, and the terms describing the
apparent quadrupole moment. The subsequent analysis then takes
advantage of the considerable simplicity possible in the odd-parity
case. A model problem is defined with a central source of odd-parity
waves which propagate outward through a perfect-fluid galaxy. In
Sec.~IV a Green function solution to the odd-parity gravitational wave
problem is constructed which shows clearly the relationships among the
source strength of the waves, the intensity of the outgoing radiation,
and the various terms that can be identified as the quadrupole moment.
Section V takes up the problem of the extent to which the galaxy is
transparent to (odd-parity) radiation.  Numerical results are then
presented which show that even for strong gravitational fields, the
effect of gravitational potential wells on the propagation of high
frequency radiation is negligible (except, of course, for the well
known redshift effect). A summary and discussion of conclusions is
given in Sec.~VI.

\section{The Kundu suppression and the KNR model}

Kundu's arguments are framed in the Newman-Penrose\cite{NP} (hereafter NP)
formalism
and are based on the Weyl projection  $\Psi_0$ in that formalism. For an
outgoing solution
$\Psi_0$ takes the form
\be
\Psi_0 = \psi^0_0(u,\theta,\phi)r^{-5}+{\cal O}(r^{-6})
\ee
where $u$ is retarded time.  Due to its $r^{-5}$ fall off at large
$r$, the quantity $\Psi_0$ is  not usually viewed as a direct measure
of radiation intensity for outgoing solution, but rather as encoding
information about the multipole moments of the source in the near zone
(i.e., at distances from the source small compared to a wavelength).
The shear
\be
\sigma=\sigma_0(u,\theta,\phi)r^{-2}+{\cal O}(r^{-4})
\ee
is well accepted as carrying the information about gravitational
wave energy density, specifically in the Bondi news function\cite{BvM}
$d\sigma_0/du$.

Kundu considers linear perturbations about a Schwarzschild background of mass
$M$
and shows that there is a simple relationship between the quantity
$\psi^0_0$ that carries information about multipole moments, and the
quantity $\sigma_0$ that carries information about radiation.
To express this relationship it is convenient to define the ``despun''
\cite{PriceII} equivalents
$\hat{\Psi}_0$ and $\hat{\sigma}$, of the spin-weight +2 quantities
$\Psi_0$ and $\sigma$, by
\be
\hat{\Psi}_0\equiv(1/2)\bar{\partial}\bar{\partial}\Psi_0
\ \ \ \ \hat{\sigma}\equiv(1/2)\bar{\partial}\bar{\partial}\sigma\ ,
\ee
where, on spin-weight +2 quantities,
\be
\bar{\partial}\bar{\partial}\equiv\left(
\frac{\partial}{\partial\theta}+\cot{\theta}-\frac{i}{\sin{\theta}}\frac{\partial}{\partial\phi}
\right)
\left(
\frac{\partial}{\partial\theta}+2\cot{\theta}-\frac{i}{\sin{\theta}}\frac{\partial}{\partial\phi}
\right)\ .
\ee
For a multipole mode, of index $\ell$, in terms of despun quantities, we find
\be\label{rawKundu}
\frac{d^3\hat{\psi^0_0}}{du^3}=
-\frac{1}{4}\frac{(\ell+2)!}{(\ell-2)!}\frac{d\bar{\hat{\sigma_0}}}{du}
-3M\frac{d^2\hat{\sigma_0}}{du^2}\ ,
\ee
in which the bar over $\hat{\sigma_0}$, in the first term on the right,
indicates
complex conjugation.
A useful feature of despun quantities is that their real and imaginary
parts correspond respectively to even- and odd-parity modes, so that, for
$\ell=2$, we have
from the real and imaginary parts of (\ref{rawKundu})
\be
\frac{d^3\left( \hat{\psi^0_0} \right)_{\rm even}}{du^3}=-\frac{6}{r^2}\left[
\frac{d\left( \hat{\sigma_0} \right)_{\rm even}}{du}
+\frac{M}{2}\frac{d^2\left( \hat{\sigma_0} \right)_{\rm even}}{du^2}
\right]
\ee\be
\frac{d^3\left( \hat{\psi^0_0} \right)_{\rm odd}}{du^3}=\frac{6}{r^2}\left[
\frac{d\left( \hat{\sigma_0} \right)_{\rm odd}}{du}
-\frac{M}{2}\frac{d^2\left( \hat{\sigma_0} \right)_{\rm odd}}{du^2}
\right]\ .
\ee
When time dependence $e^{i\omega t}$ is assumed, the result becomes
\be
\hat{\sigma_0}=\pm\frac{\omega^2}{6(1\pm i\omega M/2)}\hat{\psi^0_0}\ ,
\ee
with the + signs applying for even parity perturbations, and the - signs for
odd.

Kundu interprets this equation as telling us that the radiation
amplitude, for a quadrupole source, is reduced due to the mass of the
Schwarzschild background by the factor $(1 \pm i\omega M/2)^{-1}$, so
that the radiation power flux (proportional to $|d\sigma_0/du|^2)$ is
reduced by his suppression factor $(1+\omega^2M^2/4)^{-1}$.

Kundu's arguments depend crucially on his interpretation of
$\psi_0^0$ as the quadrupole moment of the source (aside from multiplicative
factors).
There are two types of justification given by Kundu for this
identification.  First, he argues\cite{Kundu} that this identification
is valid in linearized theory\cite{JN}, and is valid in the full
nonlinear theory\cite{Kundu88} for stationary spacetimes.
Furthermore, in the time dependent case $\psi_0^0$ has the required
transformation behavior for the quadrupole moment. As a second and
distinct justification, Kundu considers a gravitational wave source in
a massive galaxy and argues that the source integral for the
quadrupole moment will be affected by the galaxy only to order $M/R$,
and therefore the quadrupole moment will be negligibly different from
that if the source were in flat spacetime.

A model problem has recently been published which suggests that
Kundu's suppression factor is a mathematical artifact, and not of
physical importance. Kozameh, Newman and Rovelli\cite{KNR} (hereafter
KNR) use a simple model of a perturbative scalar field $\Phi$, and
show that the same apparent suppression applies to quadrupole
radiation in this model as in the case of gravitational waves.  They
consider the equations for $\ell=2$ scalar perturbations in a
Schwarzschild background of mass $M$ and define $\psi$ with
$\Phi=\psi(r,u) Y_{2m}(\theta,\phi)$, where $r$ is the Schwarzschild
radial coordinate, $u$ is retarded time, and $Y_{2m}$ is a standard
(spin-weight 0) $\ell=2$ spherical harmonic. An outgoing solution will
then have the form
\be
\psi=\frac{\psi_0}{r}
+\frac{\psi_1}{r^2}+\frac{\psi_2}{r^3}+\frac{\psi_3}{r^4}+{\cal O}(r^{-5})\ ,
\ee
in which the $\psi_k$ are functions only of $u$.  For a stationary
solution $\psi_0$ and $\psi_1$ would vanish, and $\psi_2$ would be the
quadrupole moment. To emphasize that the situation may be more
ambiguous for dynamical solutions in curved spacetime, KNR refer to
$\psi_2$ as the ``field quadrupole,'' and designate it by $Q^f$. The
equations for the scalar field in the Schwarzschild background then
show that $\psi_0$, the radiative part of $\psi$ is related to $Q^f$ by
\be
\psi_0=-\frac{\omega^2}{3}\ \frac{Q^f}{1+i\omega M/6}.
\ee
For a given time changing quadrupole, therefore, the radiated power is
reduced by the factor $(1+\omega^2M^2/36)^{-1}$ from what it would be
in flat spacetime. KNR assume that this suppression is analogous to
the suppression of gravitational radiation found by Kundu, and that
the scalar example provides a simple model for understanding the
suppression. KNR then proceed to use the scalar model to investigate
the question whether the field quadrupole $Q^f$ is really what is
usually considered the ``quadrupole moment'' of a gravitational wave
source, that is, whether $Q^f$ is the same as the ``source quadrupole
moment'' $Q$.

To address this question they consider a very simple model: a source
inside a massive spherical shell of radius $R$. Since the spacetime
inside the shell is flat, $\psi$ has a simple closed-form outgoing
solution for $r<R$. In the interior of the shell $Q^f$ can easily be shown to
be
identical to the source quadrupole $Q$. The interior and exterior
solutions are then matched at $r=R$ with the condition that the scalar
field is continuous across the shell.  A consequence of this is to
require that $Q^f$ is {\em discontinuous} across the shell; it
increases across the shell by the factor $1+i\omega M/6$. This
enhancement factor cancels the suppression factor and one concludes
that the relationship between the scalar radiation, and the scalar
source quadrupole (aside from negligible factors of order $M/R$) is
the same as in flat spacetime.

The KNR model is very suggestive of the root of the problem, that the
(exterior) ``field quadrupole'' differs from the source quadrupole.
But one might ask what details of the model could be changed to make
the argument more convincing.  Two details would seem to deserve the
most attention.  First, the model involves scalar fields and it is
difficult to be certain that the lesson of scalar fields applies to
gravitational perturbations.  A second detail of the KNR model is more
important. In the KNR model the assumption that $\psi$ is continuous
is tantamount to assuming that the shell is transparent to scalar
radiation. But for a shell which cannot absorb or reflect radiation,
in a time invariant background, we know {\em a priori} that the
radiation outside must be related to the source in the same way as the
radiation inside.  The matching condition, then, eliminates at the
outset any possibility of a Kundu effect.  A more convincing
calculation would model the interaction of the waves and galaxy to
allow for interactions, in particular for backscatter.

In the following sections we attempt to fill in some of these details.
We study gravitational waves produced by a central source and
propagating outward through a ``galaxy.''  We include all effects of
interaction and show that a nearly Newtonian galaxy is indeed
transparent to the propagation of waves.  We also show explicitly why
the $r^{-2}$ term in the outgoing solution {\em does} correspond to
the quadrupole moment near the source at the center of the galaxy, but
not outside the galaxy.

\section{Radiation suppression in the Zerilli \\ and Regge-Wheeler equations}

Although Kundu's analysis of gravitational radiation is carried out in
the NP formalism, it turns out to be convenient, as well as
instructive, to look at the problem in terms of metric perturbations.
We start by showing the relation of  the suppression
factor in the two formalisms.
Both inside and outside the galaxy we take the form of the background metric to
be
\be
\label{genmet1}
ds^2=-e^\nu dt^2+e^\lambda dr^2+r^2(d\theta^2+\sin^2\theta d\phi^2),
\ee
with $\nu$ and $\lambda$ functions of $r$ only.
We define the radial variable $r_*$  by
\be
dr/dr_*\equiv e^{(\nu-\lambda)/2}\equiv e^{\alpha(r)},
\ee
and the retarded time $u$  by
\be
u\equiv t-r_*.
\ee

In the Schwarzschild geometry, even-parity perturbations for a
particular multipole moment (with $\ell\ge2$) are conveniently
described by the Zerilli\cite{Zerilli} function $Z^{(+)}$ which
satisfies a simple potential type equation.
\be
\left( \frac{\partial^2}{\partial r_*^2}-\frac{\partial^2}{\partial t^2}
\right)Z^{(+)}=V^{+}Z^{(+)},
\ee
where, for $\ell=2$,
\be
V^{(+)}=\frac{(1-2M/r)}{r^2(1+3M/2r)^2}\left[ 6+\frac{6M}{r}
+\frac{9M^2}{r^2}+\frac{9M^3}{2r^3}
\right]\ .
\ee
(We use here the notation of  Eq.~(62), Sec. 24, of
Chandrasekhar\cite{Chandra}.) From this Zerilli equation one infers
that for outgoing radiation $Z^{(+)}$ has the form
\be
\label{Zz}
Z^{(+)}=z_0^{(+)}(u)+z_1^{(+)}(u)r^{-1}+z_2^{(+)}(u)r^{-2}+z_3^{(+)}(u)r^{-1}\cdots,
\ee
and that, for $\ell=2$,
\bd
dz_1^{(+)}/du=3z_0^{(+)}
\ed\be
\label{Z2}
dz_2^{(+)}/du=z_1^{(+)}-3Mz_0^{(+)}
\ee
\bd
dz_3^{(+)}/du=-Mz_1^{(+)}+(21/4)M^2z_0^{(+)}.
\ed
These equations show that $z_0^{(+)}$ and $z_1^{(+)}$ vanish for
stationary $\ell=2$ perturbations. For stationary solutions the
$z_2^{(+)}$ term is the first nonvanishing term, and is considered to
carry information about the source quadrupole moment. For
nonstationary solutions it is the function $z_0^{(+)}$ that carries
information about gravitational radiation, since the gravitational
wave power is proportional to $|dz_0^{(+)}/du|^2$. From Eqs.~\ref{Z2}
we can then deduce a relationship,
$d^2z_2^{(+)}/du^2=3(z_0^{(+)}-Mdz_0^{(+)}/du)$, between the radiation
quantity $z_0^{(+)}$, and the ``quadrupole moment'' $z_2^{(+)}$.  This
relationship implies a suppression factor $|1-i\omega M|^{-2}$, which
is different from the factor found by Kundu. The difference arises
from the difference in the choice of the ``quadrupole moment'' one
infers from $\Psi_0$ and from $Z^{(+)}$, and demonstrates the
importance of that choice in the inference of suppression of
radiation.

For even-parity perturbations, the relationship between Re$(\Psi_0)$ and
$Z^{(+)}$ is given in Sec.~31, Eq.~(352) of Chandrasekhar\cite{Chandra} as
\begin{displaymath}
-\frac{{\rm Re}(\Psi_0)\sin^2\theta}{C^{-3/2}_{\ell+2}}
\left(1-\frac{2M}{r}\right)
=\frac{3+3M/r+9M^2/2r^2+9M^3/4r^3}{r^3(1+3M/2r)^2}Z^{(+)}
\end{displaymath}
\be
\label{ZnPsi}
+\left[
\frac{1}{r}\frac{\partial}{\partial u}+\frac{1-3M/r-3M^2/2r^2}{r^2(1+3M/2r)}
\right]\frac{\partial}{\partial r}Z^{(+)}\ .
\ee
The following should be noted about our use of that result here: (i)
The Zerilli functions defined by different authors differ by
multiplicative factors, but overall multiplicative factors will not
affect the frequency dependent suppression factor. (ii) The Weyl
projection $\Psi_0$ is invariant with respect to infinitesimal tetrad
rotations and infinitesimal coordinate changes. We therefore need not
be concerned, for example, that Chandrasekhar employs a nonstandard
coordinate gauge.  (iii) Chandrasekhar assumes azimuthal symmetry for
$\Psi_{0}$ and angular dependence
$C_{\ell+2}^{-3/2}(\theta)\csc^2\theta$, which is proportional to the
spin-weight 2 spherical harmonic $\ _2Y_{\ell m}$, for $m=0$. For this
case the angular functions are pure real, and the real and imaginary
parts of $\Psi_0$ describe, respectively, even- and odd-parity
perturbations.

When the outgoing form for $Z^{(+)}$ in Eq.~\ref{Zz} is substituted in the
righthand side of Eq.~\ref{ZnPsi}, and  Eqs.~\ref{Z2}
are used, we find
\be
-\frac{{\rm Re}(\Psi_0)\sin^2\theta}{C^{-3/2}_{\ell+2}}
=\frac{q(u)}{r^5}+{\cal O}(r^{6}).
\ee
Here the quadrupole moment $q(u)$
is given by
\be
\label{defq}
q(u)\equiv z_2^{(+)}(u)+(3M/2)z_1^{(+)}(u).
\ee
Aside from numerical multiplicative factors, $q(u)$ is equal to
$\psi^0_0$, and is what Kundu interprets as the source quadrupole
moment. We note that Eqs.~\ref{Z2} and \ref{defq} give us
$d^2q/du^2=3[z_0^{(+)}+(M/2)dz_0^{(+)}/du]$, and hence the even-parity Kundu
suppression factor $(1+i\omega M/2)^{-1}$.

The function $Z^{(-)}$ (in the notation of Chandrasekhar), which
describes odd-parity metric perturbations in the Schwarzschild
geometry, satisfies the potential type
``Regge-Wheeler''\cite{RW} equation
\be
\label{RWE}
\left( \frac{\partial^2}{\partial r_*^2}-\frac{\partial^2}{\partial t^2}
\right)Z^{(-)}=V^{(-)}Z^{(-)},
\ee
where
\be
V^{(-)}=\frac{(1-2M/r)}{r^2}\left[\ell(\ell+1)-6M/r \right].
\ee
For a solution of the form
\be
Z^{(-)}=z_0^{(-)} + z_1^{(-)}/r+z_2^{(-)}/r^2+z_3^{(-)}/r^3+\cdots,
\ee
the Regge-Wheeler equation tells us, for $\ell=2$ multipoles, that
\bd
dz_1^{(-)}/du=3z_0^{(-)}
\ed
\be
\label{z-2}
dz_2^{(-)}/du=z_1^{(-)}-(3M/2)z_0^{(-)}
\ee
\bd
dz_3^{(-)}/du=0.
\ed

The relationship of $\Psi_0$ and $Z^{(-)}$, for odd-parity
perturbations, is given in Sec. 31, Eq. (345) of Chandrasekhar\cite{Chandra} as
\bd
-\frac{{\rm Im}(\partial\Psi_0/\partial u)\sin^2\theta}{C^{-3/2}_{\ell+2}}
\left(1-\frac{2M}{r}\right)
=\frac{1}{2r^3}\left[ \ell(\ell+1)-\frac{6M}{r} \right]Z^{(-)}
\ed
\be\label{reln}
+\left[\frac{1}{r^2}\left(1-\frac{3M}{r}+
r\frac{\partial}{\partial u}  \right) \right]\frac{\partial}{\partial
r}Z^{(-)}\ .
\ee
When the expansion in (\ref{z-2}), for an outgoing $\ell=2$ mode, is put into
(\ref{reln})
we find
\be
-\frac{{\rm Im}(d\Psi_0/du)\sin^2\theta}{C^{-3/2}_{\ell+2}}
\left(1-\frac{2M}{r}\right)
= \frac{z_2^{(-)}}{r^5} +{\cal O}(r^{-6}),
\ee
so that, aside from multiplicative constants, $z_2^{(-)}$ is equal to
$d\psi_0^0/du$ and is the derivative of what Kundu identifies as the
quadrupole moment. From Eqs.~(\ref{z-2})
we have that $d^2z_2^{(-)}/du^2=3[z_0^{(-)}-(M/2)dz_0^{(-)}/du]$
and hence the odd-parity Kundu suppression factor $(1-i\omega M/2)$.

The mathematics of the odd-parity modes can be much simpler than that
for even-parity modes since the former modes do not couple to the
perturbations of a perfect fluid. The issue of suppression is the
same for both parities, so we choose to take advantage of the opportunity
for simplicity and we consider below only odd-parity perturbations.

In the standard formalism for metric perturbations, odd-parity motions
are described as deviations of the metric in (\ref{genmet1}). We follow
here the notation of Thorne and Campolattaro\cite{TC}, in which the
Regge-Wheeler\cite{RW} gauge is used and azimuthal symmetry is assumed.
For odd-parity perturbations of multipole index $\ell$, the
only nonvanishing metric perturbations are, in this notation,
\be
\delta g_{t\phi}=h_0(r,t) \sin\theta \partial P_\ell(\cos\theta)/\partial\theta
\ee
\be
\delta g_{r\phi}=h_1(r,t) \sin\theta \partial
P_\ell(\cos\theta)/\partial\theta\ ,
\ee
where $P_\ell$ indicates the Legendre polynomial of index $\ell$.
In terms of the notation of Thorne and Campolattaro, the Chandrasekhar
function $Z^{(-)}$ is
\be
\label{Chandra-TC}
Z^{(-)}=e^\alpha h_1/r\ .
\ee
The Schwarzschild perturbation function $Z^{(-)}$ is, of course, only
defined in the exterior vacuum of the galaxy, where
$e^\alpha=e^\nu=e^{-\lambda}=1-2M/r$, but (\ref{Chandra-TC}) allows us
to extend the definition of $Z^{(-)}$ to the interior. From the
odd-parity field equations given by Thorne and Campolattaro,
$Z^{(-)}$, in the interior and exterior, is found to obey the
following generalization of (\ref{RWE})
\be
\label{intRWE}
\left( \frac{\partial^2}{\partial r_*^2}-\frac{\partial^2}{\partial t^2}
\right)Z^{(-)} -\frac{1}{r^2}\left[
e^\nu\ell(\ell+1)- 3r\frac{d\alpha}{dr}e^{2\alpha}
\right]Z^{(-)}
={\cal S}.
\ee
Here the source term ${\cal S}$ is defined, in terms of the
perturbations $\delta R_{\theta\phi}$ and $\delta R_{r\phi}$ of the Ricci
tensor, by
\be\label{sourcedef}
2re^\alpha\left[\frac{\partial}{\partial r}\left( \frac{e^\nu\delta
R_{\theta\phi}}{r^2}  \right)
-\frac{e^\nu\sin^2\theta}{r^2}\frac{\partial}{\partial\theta}\left(\frac{\delta
R_{r\phi}}{\sin^2\theta} \right)
\right]
=-{\cal S}\sin^2\theta\frac{\partial}{\partial\theta}\left[\frac{1}{\sin\theta}
\frac{\partial}{\partial\theta}P_\ell(\cos\theta)   \right]\ .
\ee

We apply (\ref{intRWE}) to the following model. At the center of a
massive galaxy, of mass $M$ and radius $R$ there is a source of
odd-parity gravitational waves at frequency $\omega$.  The source is
confined to the region for $r$ less than some source radius $r_S$. The
source, e.g., an oscillating neutron star, must
of course not consist of a perfect fluid, since odd-parity
perturbations do not couple to the motions of perfect fluids. For this
reason we take the matter of the galaxy to be a perfect fluid, so that
there can be a clean separation between the generation and the
propagation of the gravitational waves, a separation that is not
possible for even-parity waves.

It is worth noting here that the resulting mathematical formulation
differs very little from that for a scalar field of the type
considered by KNR. Let scalar field $\Phi$ have a source density
$\Sigma$ so that
\be
\Phi_{,\mu}^{;\mu}=\Sigma\ .
\ee
In the spacetime of (\ref{genmet1}), for a multipole of index $\ell$, this
equation reads
\be
\left( \frac{\partial^2}{\partial r_*^2}-\frac{\partial^2}{\partial t^2}
\right)\left(r\Phi\right) -\frac{1}{r^2}\left[
e^\nu\ell(\ell+1)+e^{2\alpha}r\frac{d\alpha}{dr}
\right]\left(r\Phi\right)
=re^\nu\Sigma\ .
\ee
The difference between the form of this equation for $r\Phi$ and
(\ref{intRWE}) for $Z^{(-)}$ is only in the details of terms of order
$M/R$.  We will show, in the next section, that these terms affect
detailed numerical results but, for a nearly Newtonian galaxy, cannot
cause significant suppression of radiation. Other features of the
scalar and the odd-parity problems are parallel. In particular, for
both cases we can consider a compact central source (no radiation
originating from the bulk of the galaxy) and the matching conditions
at the surface of the galaxy are that the fields and their radial
derivatives are continuous.

\section{Analysis  of outgoing waves}

To investigate the nature of outgoing solutions we take the time
dependence of the source, and of $Z^{(-)}$ to be $e^{i\omega t}$, and we write
\bd
Z^{(-)}=\psi(r) e^{i\omega u}\ \ \ \ {\cal S}=e^{i\omega t}S(r)\ .
\ed
The equation for odd-parity waves then takes the form
\bd
\psi''+(\alpha'-2i\omega e^{-\alpha})\psi'
-\frac{1}{r^2}[e^\lambda \ell( \ell+1)-3r\alpha'
]\psi
\ed\be
\label{simplepsi}
=e^{-2\alpha}e^{i\omega r_*} S(r) ,
\ee
where $'$ denotes differentiation.

It is straightforward, in principle, to construct a Green function
solution to Eq.~(\ref{simplepsi}) from two homogeneous solutions.  We
define a ``central" solution, $\psi_c$, as the homogeneous solution
which is well behaved at $r\rightarrow0$, with the limit
\be
\psi_c(r)\stackrel{r\rightarrow0}{\longrightarrow} r^{\ell+1}.
\ee
The second solution is taken to be the ``wave" solution $\psi_{\rm w}$ defined
by the condition
that it represents outgoing waves at large radii. The mathematical condition
on this asymptotically outgoing solution is
\be
\label{asym}
\psi_{\rm w}(r)\stackrel{r\rightarrow\infty}{\longrightarrow}
1+{\cal O}(1/\omega r) .
\ee
We define $W\equiv W(\psi_c,\psi_{\rm
w})=\psi'_{\rm w}\psi_c-\psi'_c\psi_{\rm w}$ to be the Wronskian of these
two solutions, and we note that $W$ must have the form
\be\label{Wronsk}
W=e^{2i\omega r_*}e^{-\alpha(r)}/K
\ee
in which $K$ is a constant.

If the source is confined to the region inside some radius $r_S$, then for
$r>r_S$ the Green function solution takes the form
\begin{displaymath}
\psi(r)=K\psi_{\rm w}(r)\int_0^{r_S}
\psi_c(r)S(r)e^{-\alpha}e^{-i\omega r_*}\, dr
\end{displaymath}\be
\label{GFS}
\equiv K\psi_{\rm w}(r)I_\omega,
\ee
For $r_{S}\ll R$, and in the long wavelength $(\omega r_S\ll 1)$
limit, the source integral $I_\omega$ has the approximate value
\begin{displaymath}
I_\omega\approx e^{-\nu_0/2}\int_0^{r_S}r^{\ell+1}S(r) dr.
\end{displaymath}
Note that the absence of a conical singularity requires
$\lambda\rightarrow0$ at $r\rightarrow0$, but $\nu(r=0)\equiv\nu_0$ will
in general be of order $M/R$. It should also be noted that, aside from
multiplicative numerical constants, the integral above is the usual integral
for the
$\ell^{\rm th}$ multipole moment of the source.

The function $\psi_{\rm w}(r)$ corresponds to the solution that is
asymptotically outgoing, but, due to backscatter, at small radius
($r\ll R$), it does not in general have the appearance of a {\em
locally} outgoing solution. We define a locally outgoing solution by
the high frequency expansion
\be\label{psioutexp}
\psi_{\rm out}(r) =1+i\frac{a(r)}{\omega_0}
+\frac{b(r)}{\omega_0^2}+i\frac{c(r)}{\omega_0^3}+\cdots.
\ee
Here
\bd
\omega_0\equiv\omega e^{-\nu_0/2}
\ed
is the blueshifted frequency in the central region of the galaxy. This
frequency governs the wavelength ($\lambda=2\pi c/\omega_0$ for $r\ll
R$) in the central region and is therefore the appropriate parameter
to simplify (\ref{psioutexp}).  By solving the homogeneous wave
equation (\ref{simplepsi}) to various orders in $\omega_0$ we find,
for example, that $a(r)$ and $b(r)$ must satisfy
\be
\label{aprime}
a'=\frac{1}{2}e^\alpha e^{-\nu_0/2}\left[  e^\lambda\frac{\ell(\ell+1)}{r^2}
-3\frac{\alpha'}{r}
      \right]
\ee
\be
b'=\frac{1}{2}e^\alpha e^{-\nu_0/2}\left[a''+a'\alpha' -
a\left\{e^\lambda\frac{\ell(\ell+1)}{r^2} -3\frac{\alpha'}{r}
\right\}\right] .
\ee

The functions $a, b, c, d, \ldots$ are determined only after the metric
functions $\nu, \lambda$ are specified,  but we can state some general
conclusions. For $r\ll R$ the metric coefficients can be expanded in
powers of r and, for a geometry nonsingular at $r=0$, we have $\nu'=0$
and $\lambda'=0$ at $r=0$.
As a result, the solutions for $a, b, \ldots$ take the form
\be\label{aetc}
a=-\frac{1}{2r}\frac{(\ell+1)!}{(\ell-1)!}+\frac{M}{R^2}\left[
a_{1}\frac{r}{R}  +a_{2}\frac{r^2}{R^2}+\cdots
\right]
\ee
\be\label{betc}
b=-\frac{1}{8r^2}\frac{(\ell+2)!}{(\ell-2)!}+\frac{M}{R^3}\left[
b_0 \ln\left( \frac{r}{R}\right) +b_{1}\frac{r}{R}  +\cdots
\right]
\ee
\be
c=\frac{1}{48r^3}\frac{(\ell+3)!}{(\ell-3)!}+\frac{M}{R^4}\left[
\tilde{c}_{-1}\frac{R}{r}\ln\left(\frac{r}{R}\right)
+c_{-1}\frac{R}{r}   +\cdots
\right] ,
\ee
and so forth. Here the coefficients $a_k, b_k, \ldots $ are
numerical constants aside from corrections of order $M/R$. More
precisely they are functions of the parameters of the interior geometry which
have finite limits as $M/R\rightarrow0$.

In the original homogeneous equation, Eq.~(\ref{RWE}) with the source set to
zero, there is
symmetry with respect to $t\rightarrow-t$ and complex conjugation. From
this symmetry we get a second, ingoing, solution
\be
\psi_{\rm in}=e^{2i\omega r_*}\bar{\psi}_{\rm out}\ ,
\ee
in which the bar denotes complex conjugation.

The asymptotically outgoing solution $\psi_{\rm w}(r)$ must be some
combination of $\psi_{\rm out}$ and $\psi_{\rm in}$, which  we write as
\be\label{inout}
\psi_{\rm w}(r)={\cal T} \psi_{\rm out}(r) +{\cal R} \psi_{\rm in}(r) ,
\ee
where the constants ${\cal T}$ and ${\cal R}$ can be considered
transmission and reflection coefficients. The value of $|{\cal T}|^2-|{\cal
R}|^2$
is computed by considering the Wronskian of $Z_{\rm w}^{(-)}
\equiv e^{i\omega u}\psi_{\rm w}$ and its complex conjugate,
and is found to be equal to unity aside from small corrections.
(The value of $|{\cal T}|^2-|{\cal R}|^2$ can be made precisely unity by a
small correction
 in the normalization of $\psi_{\rm w}$.)

For $r\ll R$ the solution
$\psi_{\rm w}(r)$ in (\ref{inout}) can be expanded in powers of $r$. For
$\ell=2$ this  gives
\bd
\psi_{\rm w}={\cal T}\left\{
1+\frac{i}{\omega_0}\left[ -\frac{3}{r}+{\cal O}(Mr/R^3)
\right]
+\frac{1}{\omega_0^2}\left[-\frac{3}{r^2} +{\cal
O}\left(\frac{M}{R^3}\ln\left(\frac{r}{R} \right)\right) \right]\right.
\ed
\be\label{psiwexp}
\left. +\frac{i}{\omega_0^3}  {\cal O}\left(\frac{M}{R^3r}\ln\frac{r}{R}\right)
+\cdots\right\}
\ee
\bd
+{\cal R}e^{2i\omega r_*}\left\{ 1-\frac{i}{\omega_0}\left[ -\frac{3}{r}+{\cal
O}(Mr/R^3)
\right]
+\frac{1}{\omega_0^2}\left[-\frac{3}{r^2} +{\cal
O}\left(\frac{M}{R^3}\ln\left(\frac{r}{R} \right)\right)  \right]\right.
\ed
\bd
\left. - \frac{i}{\omega_0^3}  {\cal
O}\left(\frac{M}{R^3r}\ln\frac{r}{R}\right)
+\cdots\right\}
\ed
For $r\ll R$ then, the term in $\psi_{\rm w}$ that goes as $r^{-2}$ is
\be
\frac{{\cal T}+{\cal R}}{\omega_0^2}\left\{-3+{\cal O}\left(\frac{M}{\omega^2
R^3} \right)  \right\}
\ee
Thus, aside from corrections which are small for  high frequency ($\omega
R\gg1$) sources,
the $r^{-2}$ term in $\psi$ is
\be
({\cal T}+{\cal R})(-3/\omega_0^2)KI_\omega .
\ee
 If, as expected, backscatter is insignificant, then $|{\cal T}|\approx 1$ and
$|{\cal R}|\ll 1$ so that the $r^{-2}$ term is approximately
\be
(-3/\omega_0^2)KI_\omega .
\ee
The $r^{-2}$ term therefore gives a direct measure of the quadrupole source
integral. From Eqs.~(\ref{asym}) and (\ref{GFS}) it follows that
$\psi\rightarrow KI_\omega$ as $r\rightarrow\infty$, and that the $r^{-2}$
term in the deep interior  also gives a measure of the intensity of
the outgoing radiation.

For the exterior solution, very different conclusions follow. Here it
is possible to expand $\psi_{\rm w}$ in inverse powers of $r$ as
\begin{displaymath}
\psi_{\rm w}=1+\frac{i}{\omega}\left[
-\frac{3}{r}+A_1\frac{M}{r^2} +A_2\frac{M^2}{r^3}+\cdots\right]
\end{displaymath}
\be
\label{psiw}
+\frac{1}{\omega^2}\left[
-\frac{3}{r^2}+B_1\frac{M}{r^3} +\cdots\right]
+\frac{1}{\omega^3}\left[
C_1\frac{M}{r^4} +\cdots\right] +\cdots ,
\ee
in which the numerical constants, $A_1=3/2, A_2=0, B_1=0$, etc., are easily
evaluated from the Schwarzschild metric functions.
In the exterior solution then the $r^{-2}$ term in
$\psi$ is
\be
(-3/\omega^2 + 3iM/2\omega)KI_\omega .
\ee
and is larger than the interior $r^{-2}$ term by the (possibly large) factor
\hbox{$(1-i\omega M/2)$}. But now the $r^{-2}$ coefficient no longer gives the
intensity of the outgoing radiation, or a measure of the source integral. If
the
coefficient is used to denote (in the notation of KNR) the ``field quadrupole,"
then it must be understood that this field quadrupole is larger, by the factor
\hbox{$(1 - i\omega M/2)$}, than the quadrupole moment which measures the
source integral, which governs the intensity of outgoing radiation at
infinity, or which governs the locally outgoing radiation deep inside
the galaxy.

It is clear mathematically why the field quadrupole and the physical
quadrupole are so different: the $ia(r)/\omega_0$ term in (\ref{psiwexp})
lacks a term that goes as $r^{-2}$. From (\ref{aprime}) we see
that the presence of such a term would require a galaxy spacetime that
tends to a singularity as $r\rightarrow0$.  The absence of such a term
is why, for a high frequency source, the $r^{-2}$ term for $r\ll R$ can be
given the same meaning---that of the quadrupole moment---as in a flat
spacetime background. In (\ref{psiw}) the $ia(r)/\omega$ term {\em
does} have a $r^{-2}$ term. This is possible because the expansion in
(\ref{psiw}) cannot be extended to small $r$. But to interpret as
the quadrupole moment the $r^{-2}$ term, in some expansion for $\psi$,
is justifiable only if that expansion can be extended to small radii.
The field quadrupole is, then, a formal construct and its use as a
physical quadrupole moment is the reason that gravitational radiation
appears to be suppressed.

This insight gives the answer to an interesting question. Let us denote
the coefficient of the $r^{-2}$ term as $Q_f$, both in the deep
interior and in the exterior. When $Q_f$ is computed in the exterior
we find a different value than in the interior, and than we would find
in the absence of the galaxy. How can the ``source integral'' for
$Q_f$, when it is computed in the exterior, have large non-Newtonian
contributions from the galaxy, especially in the case that the galaxy
is nearly Newtonian?  The answer is that the ``source integral,'' both
in the exterior and in the deep interior, is the same. It is
$I_\omega$ of (\ref{GFS}). But the way in which this source
integral enters into the expression for the coefficient of the $r^{-2}$
term, and hence into the value inferred for $Q_f$,
is different in the exterior and the interior.

In the next section we consider just what the magnitude is of the
influence of the galaxy spacetime on the outgoing radiation.

\section{Reflection and transmission of outgoing waves}

We take up here the question of what the actual influence is of the
curved spacetime of the galaxy on the propagation of gravitational
waves (specifically, of odd-parity gravitational waves).  One obvious
influence, of course, is the redshift which is built into the
expressions for the radiation. The net power in terms of coordinate
time $t$ must be independent of the distance from the source. The
locally measured proper time differs from coordinate time by
$e^{\nu/2}$ and hence the locally measured power (proportional to the
square of the time derivative of $\psi$) will differ from that far
outside the source by the redshift factor $e^\nu$.

The question of other influences on the radiation is much less obvious, and
there are
 several
ways in which it can be asked. One approach is to look at
the relation of the outgoing radiation and the source as embodied in
(\ref{sourcedef}) and (\ref{GFS}).  This approach is most transparent
if the source is taken to be  compact, i.e., $r_S\ll1/\omega$ as
well as $r_S\ll R$.  In this case, the radial derivative of $\nu$ will be
smaller (by $r_S/R$) than the radial derivative of the Ricci
components, so we can approximate
\bd
2re^{3\nu_0/2}\left[\frac{d}{dr}\left(\delta R_{\hat{\theta}\hat{\phi}}\right)
-\frac{\sin\theta}{r}\frac{d}{d\theta}\left(\frac{\delta
R_{\hat{r}\hat{\phi}}}{\sin\theta} \right)
\right]
\ed\be
=-{\cal S}\sin\theta\frac{\partial}{\partial\theta}\left[\frac{1}{\sin\theta}
\frac{\partial}{\partial\theta}P_\ell(\cos\theta)   \right]\ .
\ee
Here the terms $\delta R_{\hat{\theta}\hat{\phi}}$ and $\delta
R_{\hat{r}\hat{\phi}}$ are the perturbations of the Ricci tensor
projected on an orthonormal tetrad, and are the
quantities that would be computed (e.g., for a neutron star) by a
nearby observer. We therefore write
\be
{\cal S}=e^{3\nu_0/2}{\cal S}_{\rm local}
\ee
to indicate  the relation of the source term referred to the
coordinates of (\ref{genmet1}) and the source term measured by a local
observer.

For the compact source, with corrections of order $r_S M/R^2$ and
$1/r_S \omega$ ignored, (\ref{GFS}) can be written
\be
\psi(r)=Ke^{\nu_0} I_{\rm local}\psi_{\rm w} ,
\ee
in which
\be
 I_{\rm local}=\int_0^{r_S}r^{\ell+1}{\cal S}_{\rm local}\, dr
\ee
is the source term that would be computed by a local observer.  In the
case of flat spacetime $K$ defined by (\ref{Wronsk}) is easily shown
to be $-(i)^\ell\omega^\ell[(2\ell+1)!!]^{-1}$ so that, finally, the
relation of source and field can be written as
\be
\psi(r)=-(i)^\ell\omega^\ell[(2\ell+1)!!]^{-1}I_{\rm local}\psi_{\rm w}
e^{\nu_0} \kappa_{\rm corr}
\ee
with
\be\label{kappaK}
\kappa_{\rm corr}\equiv-(2\ell+1)!!(-i)^\ell\omega^{-\ell} K.
\ee
In this equation the influence of the galaxy is contained in the factor
$e^{\nu_0}\kappa_{\rm corr}$.

There is another, rather different, way in which the influence of the
galaxy can be viewed. One can ask what the relationship is between the
outgoing radiation far outside the galaxy, and the radiation in the
deep interior of the galaxy.  In (\ref{inout}) this relationship is
contained in the constants ${\cal T}$ and ${\cal R}$, in which ${\cal
R}$ describes, approximately, the fraction of the radiation reflected
back towards the source, due to the galaxy's spacetime curvature.
Roughly speaking, the magnitude of $|{\cal R}|$, or of $|{\cal T}|-1$,
is a measure of the extent to which the galaxy is not perfectly
transparent to gravitational radiation.  It is only an approximate
measure because there is, at the outset, a limit to the precision to
which an observer can measure radiation as if in flat spacetime. The
metric for a ``flat'' coordinate system over a region of size $L$ will
deviate from the Minkowski metric by corrections of order $(L/R_c)^2$,
where $R_c$ is the spacetime radius of curvature. For the galaxy
spacetime $R_c\sim (R^3/M)^{1/2}$, so that over one wavelength there
will be metric corrections of order $M/\omega^2R^3$.  One
manifestation of this is that we have, from the Wronskian of
$\psi_{\rm w}$ and $e^{2i\omega r_*}\overline{\psi_{\rm w}}$, and the
expressions in (\ref{psiwexp}) and (\ref{psiw}), that
\be\label{fluxcons}
|{\cal T}|^2-|{\cal R}|^2=1+{\cal O}(M/\omega^2R^3).
\ee
The ${\cal O}(M/\omega^2R^3)$ correction factor is simple to compute,
once $\nu$ and $\lambda$ are specified, from the forms for $a, b,
\ldots$.
The correction factor $M/\omega^2R^3$ will, in any case, be negligible
(of order $10^{-39}$ for kilohertz waves and ordinary galaxies).

There is a close relationship between the two viewpoints above for
looking at the influence of the galaxy. The Wronskian in
(\ref{Wronsk}) can be written
\be
W(\psi_c,\psi_{\rm w}) = {\cal T}W(\psi_c,\psi_{\rm out}) + {\cal R}
W(\psi_c,\psi_{\rm in}),
\ee
and the Wronskians on the righthand side can be evaluated to give, for
$\ell=2$,
\be\label{WronskReln}
W(\psi_c,\psi_{\rm w}) =({\cal T}+{\cal R})
e^{2i\omega r_*}e^{-\alpha}15\ \omega_0^{-2} e^{\nu_0/2}
\left[1 + {\cal O} (M/\omega^2 R^3)  \right]\ .
\ee
As in (\ref{fluxcons}), the ${\cal O} (M/\omega^2 R^3)$ correction
term is easily evaluated from (\ref{psioutexp}) and the small-radius
forms of $a(r), b(r), \ldots$, once the the metric functions $\nu$ and
$\lambda$ are specified.

When we combine (\ref{WronskReln}) with (\ref{Wronsk}) and
(\ref{kappaK}), for $\ell=2$, we have
\be\label{WronskExpl}
\kappa_{\rm corr} =e^{-3\nu_0/2} ({\cal T}+{\cal R})^{-1}
\left(1 + {\cal O} (M/\omega^2 R^3)  \right)\ .
\ee
The effect of the galaxy is then contained in two types of terms.
There are terms of order $M/\omega^2R^3$ (e.g., in (\ref{WronskExpl}) and
(\ref{fluxcons})) that are ``local'' in the sense that they can be
computed from the small-radius solutions for $\psi$. The second
influence of the galaxy is through the coefficient ${\cal R}$, and is
not local.  If there is any way in which a nearly Newtonian galaxy can
have a significant influence on the propagation of high frequency
waves, it is through the possibility that $|{\cal R}|$ is not small.

That possibility can, in fact, be ruled out with a WKB argument, but
such an argument cannot easily tell us how small $|{\cal R}|$ really is. To
find this out we have numerically integrated the equation for $\psi$
starting in the exterior, at large $r$, with the expansion in
(\ref{psiwexp}). The integration to small radius was done with the
method of Gear\cite{Gear}, suitable to the stiff differential equation
for $\psi$. At the surface $r=R$, the
field equations require that $\psi$ and $\psi'$ be continuous.
The ${\cal R}$ coefficient was extracted from the
numerically computed solution $\psi_{\rm w}$, by using the flat
spacetime solutions $\psi_{\rm out}^{\rm flat}\equiv
1-3i/r\omega_0-3/r^2\omega_0^2$, and $\psi_{\rm in}^{\rm flat}\equiv
e^{2i\omega r_*}\overline{\psi_{\rm out}^{\rm flat}}$, and the
computed quantity
\be
\label{DefRindex}
{\cal R}_{\rm index}\equiv W(\psi_{\rm out}^{\rm flat},\psi_{\rm w})
/W(\psi_{\rm out}^{\rm flat},\psi_{\rm in}^{\rm flat})\ .
\ee
This can be evaluated at small $r$ with the expansion in (\ref{psioutexp}) to
give
\be\label{Rindex}
{\cal R}_{\rm index} =\left[ {\cal R}+
\frac{a_1M}{2\omega_0^2 R^3}
 e^{-2i\omega r_*} {\cal T} \right]
\left[1+ {\cal O}\left(\frac{1}{\omega_0r} \right)
+ {\cal O}\left(\frac{r}{R} \right) \right] \ ,
\ee
where $a_1$ is the coefficient defined in (\ref{aetc}). Note that
(\ref{Rindex})
does not assume $M\ll R$; it can be used for galaxies
with relativistically strong gravity.

For a constant density interior (Schwarzschild interior) the value of $a_1$
is easily shown to be 6, independent of $M/R$, and we apply (\ref{Rindex})
to the results for ${\cal R}_{\rm index}$ found with (\ref{DefRindex})
from the numerically computed values of $\psi_{\rm w}$.
In Fig. 1 we show the real part of ${\cal R}_{\rm index}$ as a
function of $r$, for the parameters $M/R=0.1$ and $\omega R=2000$, and
we compare it to the predicted expression in (\ref{Rindex}) for the
best-fit values ${\cal R}=(0.95+4.3 i) 10^{-8}$ and ${\cal T}=e^{0.2 i}$.
Numerical runs with different parameters show that $|{\cal R}|$
is proportional to $M/\omega^2R^3$ (aside from higher order
corrections in $\omega R$), so that the ${\cal R}$ and ${\cal T}$
terms on the righthand side of (\ref{Rindex}) are of the same order.

Figure 2 gives $|{\cal R}|$ as a function of $M/R$, for
different values of $\omega R$. The plots clearly indicate that
$|{\cal R}| \approx   {2 M \over \omega^2 R^3}$ as long as $M/R\ll 1$ and
$\omega
R\gg 1$. When $M/R$ is no longer small, it remains true that $|{\cal
R}|\propto\omega^{-2}R^{-2}$, but the dependence on $M/R$ must be read
from the figure. When $\omega$ is not large compared to $R^{-1}$, the
assumptions used in deriving (\ref{Rindex}) fail as does much of the
meaning of ``reflection.''

\section{Summary and Conclusions}

We have studied a configuration in which waves propagate on a
spherical static curved spacetime background. In the geometric optics
limit, the limit of infinite frequency, the only physical influence of
the background is the familiar redshift of the waves. We have found
effects for high, but finite frequency $\omega$, for propagation in a
background of mass $M$ and radius $R$. Most notably, we have found
that the radiation reaching arbitrary distances is different from that
emitted, by a fractional correction of order $M/\omega^2R^3$.  The
computed reflection coefficients in Sec.~V may be considered the first
corrections to the geometric optics limit.

Some details of Secs.~IV and V are specific to odd-parity gravitational
waves, but with very minor modifications apply also to massless,
minimally coupled, scalar fields. The generalization to even-parity
gravitational waves is not immediate. For the odd-parity, or
scalar, case the waves propagate on the curved background of the galaxy
fluid, but there is no direct matter-fluid interaction.  For
even-parity waves propagating through a perfect fluid galaxy, or for
waves of either parity propagating through a region with more complex
material properties, the matter will, in general, oscillate in response
to the passage of the wave, and will retard and absorb radiation much
as a dielectric material interacts with an electromagnetic waves.
(The computation for even-parity waves through a perfect fluid galaxy
would be relatively straightforward to carry out with formalisms in
which fluid perturbations do not explicitly appear\cite{ChFIP}, but
the problem is made difficult by its four degrees of freedom.)  These
interactions, however, can be estimated reliably\cite{Chesters}
and except for  contrived circumstances will be very small.

What do these results imply for the possibility of suppression of
gravitational radiation, as predicted by Kundu\cite{Kundu,K2}?
The use of a model calculation specific to odd-parity waves is
irrelevant. The suppression is inferred by Kundu from the external
Schwarzschild geometry, in which the mathematics of even and
odd-parity waves is essentially the same.  Our analysis is rather
specific to a particular configuration: a compact central wave source
embedded in a massive spherical background. One might ask whether the
suppression might apply to very different configurations, such as a
source near a massive black hole. It would be strange, of course, if
the suppression ---inferred only from the Schwarzschild
background---applied for one wave source and not another within that
background. Barring that possibility, the general lessons of our
spherical configurations should apply insofar as well defined questions
can be asked about suppression. In particular, for a compact source,
in the geometrical optics limit (wavelength$\ll$ all other length
scales), the effect of background curvature should be only the standard
redshift and the bending of the null geodesics (absent in the
spherically symmetric case). For finite frequency we would expect the
first corrections from the geometric optics limit to be of order
$(\omega R_c)^{-2}$, where $R_c$ is the characteristic spacetime
radius of curvature.

The conclusions based on the configuration considered in this paper
should then give the generally correct picture of the relationship of
radiation and quadrupole moment. In that picture we are able to
distinguish a number of different ``quadrupole moments'': (i) a
quadrupole moment given by an integral over the source (ii) the
interior ``field quadrupole'' inferred from the coefficient of the
$r^{-5}$ term in the NP Weyl projection $\Psi_0$ near the source (iii)
the ``field quadrupole'' from the $r^{-5}$ term far from the source of
background curvature. We have shown that the quadrupole moments of
types (i) and (ii) are the same (aside from corrections of order
$M/\omega^2R^3$) and that they govern the outgoing radiation produced
by the source, both deep within the galaxy and outside the galaxy. The
exterior ``field quadrupole,'' however, differs significantly from the
other quadrupole moments. Its interpretation as a quadrupole moment is
based on an expansion in inverse powers of $r$, and an identification
of the expansion coefficients with those of similar expressions for
$r\ll R$.  But the expansion in the exterior cannot be extended
inward, so that the expansion coefficients do not have their usual
physical meaning. In particular the ``field quadrupole'' is not the
quadrupole moment of the source, and does not govern the radiation
produced by the source.

\acknowledgments
\nonum
We wish to thank Kip Thorne for discussions of this work. Support
for the research reported here was provided by the National Science
Foundation under grant PHY-8907937.

\figure{${\cal R}_{\rm index}$ is plotted for the numerical runs (circles)
and the value given by the analytical estimate eq. (68) (curve). The
parameters of the run are $R=100$, $M=10$, $\omega=20$. The
error bars in the numerical points are of the order of the size of the
circles.}

\figure{The absolute value of ${\cal R}_{\rm index}$ plotted as a function
of $M/R$ for two values of $\omega R$. The circles are for $\omega
R=1000$ and the stars for $\omega R=2000$. The continuous curves
represent the approximation for small $M/R$ given by $|R| \approx
{2 M \over \omega^{2} R^{3}}$.}

\end{document}